# Broadening the View of Live Programmers

## Integrating a Cross-Cutting Perspective on Run-Time Behavior into a Live Programming Environment

Patrick Rein[a], Christian Flach[a], Stefan Ramson[a], Eva Krebs[a], and Robert Hirschfeld[a]

a Hasso Plattner Institute, University of Potsdam, Germany

**Abstract** Live programming provides feedback on run-time behavior by visualizing concrete values of expressions close to the source code. When using such a *local perspective* on run-time behavior, programmers have to mentally reconstruct the control flow if they want to understand the relation between observed values. As this requires complete and correct knowledge of all relevant code, this reconstruction is impractical for larger programs as well as in the case of unexpected program behavior. In turn, *cross-cutting perspectives* on run-time behavior can visualize the actual control flow directly. At the same time, cross-cutting perspectives are often difficult to navigate due to the large number of run-time events.

We propose to integrate cross-cutting perspectives into live programming environments based on local perspectives so that the two complement each other: the cross-cutting perspective provides an overview of the run-time behavior; the local perspective provides detailed feedback as well as points of interest to navigate the cross-cutting perspective. We present a cross-cutting perspective prototype in the form of a call tree browser integrated into the Babylonian/S live programming environment. In an exploratory user study, we observed that programmers found the tool useful for debugging, code comprehension, and navigation. Finally, we discuss how our prototype illustrates how the features of live programming environments may serve as the basis for other powerful dynamic development tools.

## 1 Introduction

Live programming environments promise to make programming more accessible and programmers more productive by continuously providing feedback on the dynamic behavior of a program [10, 40, 43]. Some live programming environments generate this feedback by running parts of the program with concrete, user-provided input, often in the form of examples [4, 38]. They hint at run-time behavior by displaying intermediate run-time states or results of expressions or statements [4, 13, 14, 21]. As a result, they offer a *local perspective* on the program behavior, which allows programmers to comprehend the behavior of their programs on the most detailed level and to check any invariants or hypotheses about run-time state directly within the live programming environment. To focus this feedback on relevant parts of the program, many live programming environments allow programmers to express for which expressions or statements they want to get feedback by attaching so-called "probes" [26, 38] (see Figure 1).

While providing benefits for understanding single procedures of a program in detail, this local perspective does not scale to programming more extensive, more complex programs or systems. When working on systems, programmers need to understand not only the behavior within a single, isolated procedure but also the interplay between several or many of them.

Figure 1 shows an example in which programmers will struggle to comprehend code using the local perspective of the Babylonian Programming editor, a live programming environment supporting explicit examples and probes. In this scenario, the programmers try to understand the `AtomMorph` methods `velocity:` and `randomPositionIn:maxVelocity:` in the context of `BouncingAtomsMorph>>addAtoms:`. The method `addAtoms:` contains the example "basic example" that is the start of the execution whose intermediate states are displayed in the probe in the bottom editor. When looking at the probe in the bottom editor, the programmers discover a pattern in the values: there is always one point `10@10` followed by two seemingly random points. As there is only one call of `velocity:` in `randomPositionIn:maxVelocity:` this pattern comes as a surprise to them. At the same time, the live programming tools do not support them anymore to investigate this pattern any further.

When such limits of the local perspective are reached, programmers have to resort to other tools that allow them to explore the system behavior at a *cross-cutting perspective* that displays the control flow throughout the system. Regardless of whether they switch to programming tools working with static or dynamic data, they lose the relation to the examples they are currently working with. When switching to a tool working with static information, such as tools for browsing senders and implementers of methods, they have to manually reconstruct the control flow between relevant methods. In the case of another tool using dynamic information, such as a call trace or a debugger, they have to first recreate the running example in that tool, and then also have to navigate to the points of interest, which they already selected in the live programming tool. Either way, this lengthens the feedback loop and diminishes the experience of liveness while programming.

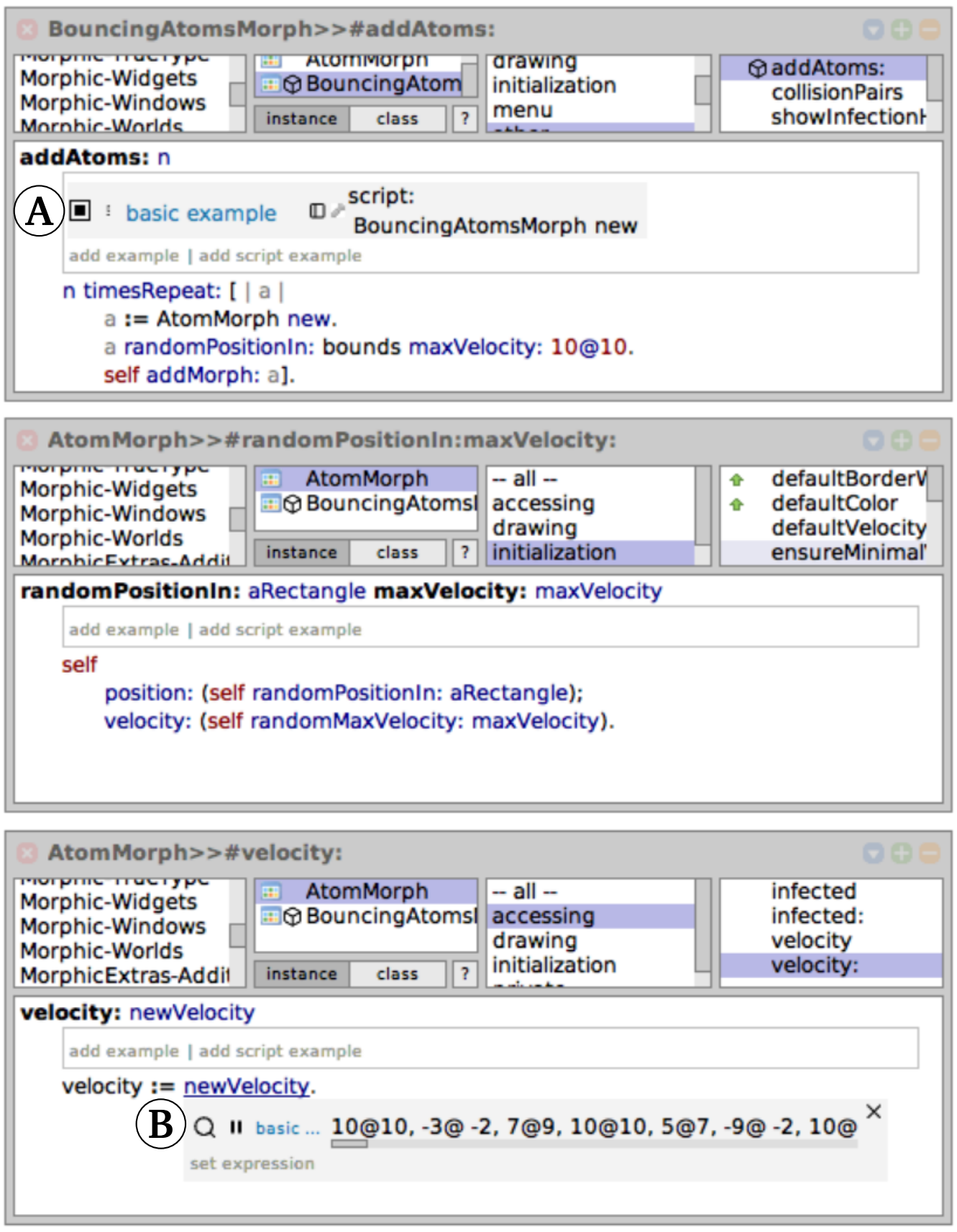

**Figure 1** Babylonian browsers showing a situation in which a local perspective will not suffice to interpret probe values. Programmers added an example called "basic example" to the method at the top (A). The values in the probe at the bottom (B) suggest at least two kinds of calls to velocity:, one of them shown in the middle. The four browser panes that are the same in (a) and (b) show packages, classes, method categories, and methods.

In this paper, we aim to extend the live programming experience to programming more extensive and complex programs. Therefore, we propose to integrate a call trace as a cross-cutting perspective on run-time behavior into a live programming environment, so that programmers can directly and without switching tools explore the interplay of methods and get live feedback on how their changes affect this interplay (see Figure 2).

However, integrating a call trace into a live programming workflow poses a new challenge, as call traces typically consist of a large number of events and are therefore difficult to navigate. To address this, we propose to regard the local perspective as the starting point for exploring the cross-cutting perspective, thereby adding a "bottom-up" navigation to the typical "top-down" perspective of call trace tools. In particular, the probes and examples placed in the local perspective already designate points of interest that programmers are aware of and are thus useful starting points. Further, to provide programmers with an initial idea of the interplay of methods, we extend the local perspective with summarized trace information [31]. Through this integration, programmers continue to benefit from short feedback cycles in the local perspective, can directly switch to the cross-cutting perspective based on concrete run-time states, and also benefit from live feedback at the cross-cutting perspective.

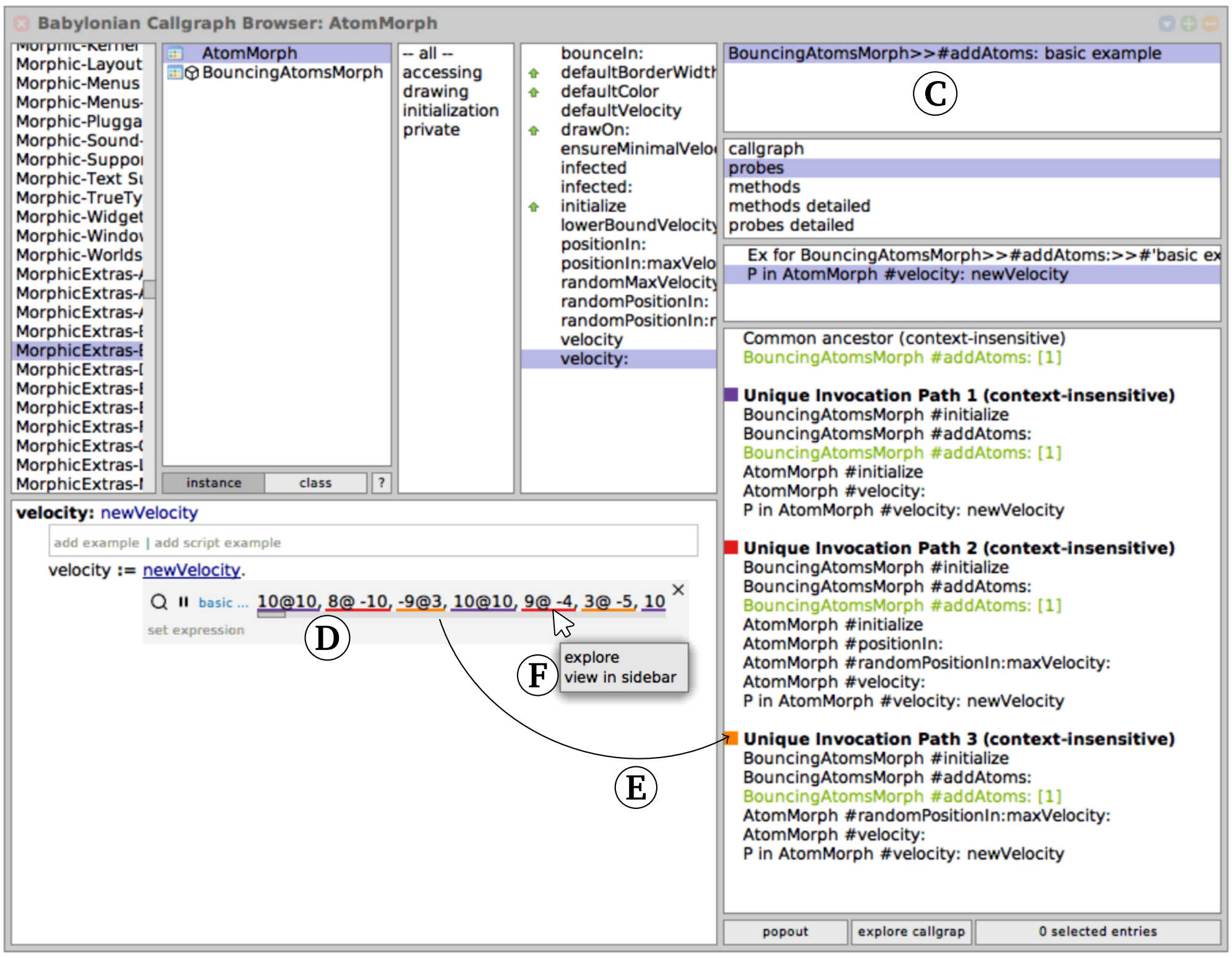

**Figure 2** A browser window in our proposed live programming environment integrating call trace views as a cross-cutting perspective on run-time behavior (C). Programmers can see that the probe values result from three different paths (D). The sidebar (C) provides further details, as it is currently showing the *summarized paths view* for the probe in velocity:. The colors of the icons correspond to the colors in the probe (E). Programmers can also jump to a trace view entry corresponding to a value (F).

The contributions of this paper are:

- The discussion of the local perspective on run-time behavior as a limiting factor for using live programming tools for working with more complex programs.
- The concept of integrating a cross-cutting perspective of run-time behavior into a live programming environment. The integration is supported by allowing programmers to navigate the cross-cutting perspective via the displayed run-time states as well as summarized traces in the local perspective.
- A prototype integrating call trace views as a basic cross-cutting perspective into the Babylonian Programming environment, an example-based live programming environment.
- An exploratory user study on the impact of the cross-cutting perspective in our prototype on dynamic tool usage.

**Outline** In the remainder of the paper, we first discuss the challenges arising from local perspectives in live programming environments and review related work and their approaches to this challenge. We then introduce our concept of integrating a cross-cutting perspective based on our prototype in the Babylonian programming environment. To examine the effects of this extension of programmer behavior, we present the results of an exploratory user study. We conclude with a discussion of the results of the study, how live programming tools can help with debugging, and how they can serve as a basis for advanced programming tools.

## 2 Local and Cross-Cutting Perspectives in Live Programming

The local perspective of most live programming environments can provide programmers with concrete and immediate feedback on the evaluation results of single statements or expressions. While this is the source of their strength, it is also the limiting factor when scaling up to programming more extensive and complex systems. To discuss this, we describe the role of local perspectives in present live programming environments and why their utility degrades in the presence of complex control flow. Further, we illustrate how present live programming environments mitigate this and to which extent they already include cross-cutting perspectives. Finally, we provide a short overview of the relevant features of the Babylonian/S environment, which serves as the foundation for our prototype.

### 2.1 The Local Perspective as a Limiting Factor

Live programming environments aim to make programming accessible and programmers more productive by providing direct feedback on the dynamic behavior of a program [40]. To provide feedback on the dynamic behavior of procedures, many live programming environments rely on user-provided input data to execute the procedures. This user-provided input serves as a mental context for programmers in which they can interpret observed values. In this paper, we call such user-provided input an *example*, as it is commonly done in example-based live programming environments [4, 38].

Feedback is provided through *always-on visualizations*, such as projection boxes in Figure 3a or probes in Figure 3b. These visualizations show the values resulting from all evaluations of an expression or statement during the execution of an example. Thereby, the visualizations are usually tied to statements or expressions. This also applies to visualizations that show the results of all statements of a procedure.

To some extent, this locality of feedback is the source of the strengths of live programming environments. The feedback is shown close to the code, which is the representation of a program that programmers are most familiar with [2]. By providing feedback close to the source code, programmers can compare the observed, concrete values more easily to the expectations they have derived from static code.

At the same time, this local perspective is also a limiting factor when working with more extensive systems. In order to make sense of the displayed values, programmers

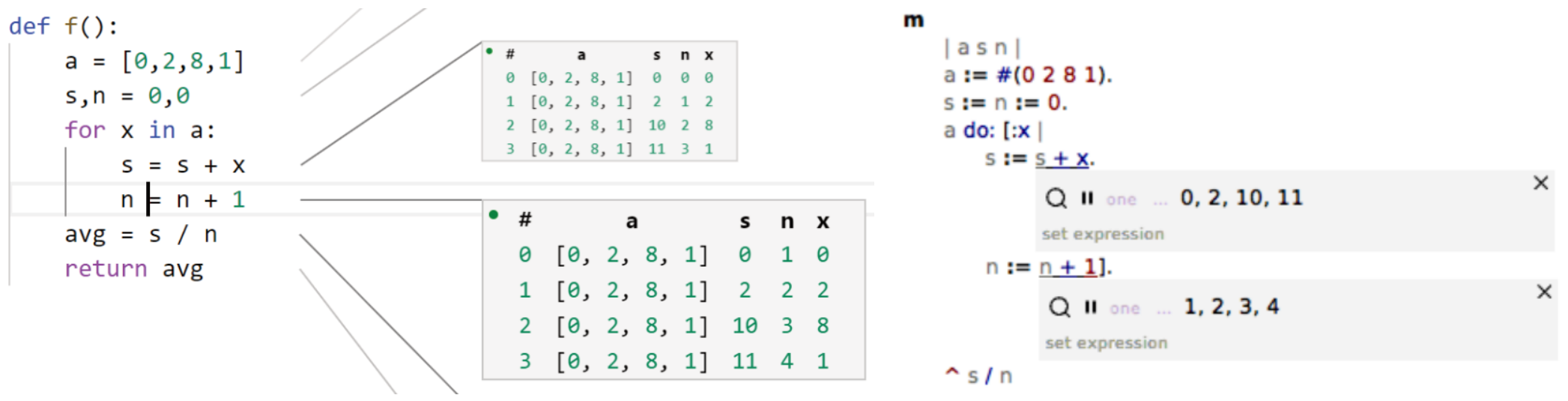

**(a)** Projection boxes. Figure from original paper [21]

**(b)** Probes in Babylonian/S [39]

**Figure 3** Two always-on visualizations representing a local perspective on run-time behavior: (a) projection boxes and (b) probes. Both provide detailed feedback on intermediate run-time state, closely tied to statements or expressions.

have to partially reconstruct the control flow of the program in their minds. For example, a method may contain two probes attached to different expressions and each probe shows one value. By only looking at the values the ordering in which the values occurred can not be deduced. The value of the first probe might have occurred before the value of the second probe, or it might have happened the other way around. To determine in which order the expressions were evaluated, programmers have to mentally reconstruct the control flow and relate it to the observed values.[1] This mental simulation is feasible for small, single procedures or whenever only limited local information is needed to determine the concrete control flow for an example.

However, when multiple or complex procedures are involved, mentally reconstructing the control flow becomes challenging and programmers have to invest considerable effort to interpret the observed values to draw conclusions about the underlying behavior. This effort required to interpret the feedback contradicts one of the goals of live programming environments, which is to offer immediate feedback on the actual behavior of the program.

This becomes especially important when the feedback in the live programming environment indicates that the program does not behave as intended. Although live programming environments are specialized for providing feedback on dynamic behavior, the limited insight into the control flow reduces the applicability of live environments for debugging tasks. Debugging involves comparing the actual control flow to the expected control flow and determining the source of erroneous state by tracing the actual control flow backwards [5, 46]. The local perspective does not make the erroneous control flow easily visible or explorable. As a result, programmers have to resort to common debugging tools, giving up on the advantages of live programming. Further, switching away from the live programming environment entails first reconstructing the example for the debugging tool and second navigating to the run-time state of interest. Finally, on changes to the program, programmers must again manually update the example and navigate to the relevant run-time state.

[1] We also call this problem the *diamond problem of live programming*, as it is unclear from the local perspective which paths were taken to reach individual values.

These manual steps lengthen the feedback loop and deteriorate the overall experience of liveness during programming.

### 2.2 Cross-Cutting Perspectives in Live Programming Environments

While a local perspective emphasizes intermediate run-time states at a confined location, a cross-cutting perspective shows the connections between different locations and run-time states. Common examples of tools providing such a perspective are control flow visualizations such as trace viewers, bundle views, or object interaction diagrams [1, 11, 20]. Other examples include visualizations of data flow such as data flow tomography [27] or changes to variables [13].

Notably, the local and cross-cutting perspectives as we define them are the two ends of a spectrum that ranges from views on a single element to views that span all elements of a program. Many views are close to the local or the cross-cutting end, for instance, probes and control-flow visualizations. However, some views also fall in between the ends such as a profiler visualization that colors each statement in a program according to its call count. This provides detailed information per statement but also connects the statements to each other.

Related live programming environments include cross-cutting perspectives to various degrees. Our proposed environment is most closely related to other live programming environments that build upon a local perspective to provide feedback.

Two environments support combinations of local and cross-cutting perspectives: Shiranui [12] and Seymour [14]. Both include a basic cross-cutting perspective by highlighting the statements that have actually been executed during the execution of an example. In addition, Shiranui allows programmers to select individual intermediate run-time states and see the dynamic slice for that value [12]. Seymour also offers a more advanced view in the form of an icicle plot [19] visualizing the stack over time. Programmers can also use the icicle plot to focus the local perspective on specific stack frames. According to the accounts of applying Seymour to student programs, the icicle plot works well for smaller programs but does not scale to more extensive programs, as it is missing relevant information such as the names of called procedures [14].

The YinYang [26] includes a cross-cutting perspective based on a custom trace. This trace displays the results of manually placed `print`-statements and allows programmers to navigate the code using them. This offers some form of cross-cutting perspective, but the manual, explicit annotation of points of interest that should appear in the trace might hamper exploration.

A different cross-cutting perspective is provided by Omnicode [13]. Instead of visualizing the control flow, Omnicode visualizes changes to the complete run-time state throughout the whole program execution. A similar perspective is part of DejaVu [15] for interactive camera-based programs, but it requires users to explicitly choose which states should be displayed over time.

Interestingly, the example-centric environment [4] does not implement a local perspective, but only a cross-cutting perspective. The program behavior is visualized in a full execution trace, that shows every single step of the execution and the concrete data used in these steps. Programmers can thus explore the full execution but also have to navigate it to determine all results of a single expression. Similarly, the Theseus [23] environment does not provide a local perspective, but a cross-cutting perspective in the form of a call tree of JavaScript methods.

Other environments focus on different design aspects and do not yet provide cross-cutting perspectives, such as Live Literals [42], Live Brackets [17], Hazel [29], and Light Table [9].

Projection Boxes [21] as a generalized visualization framework for a local perspective does not yet offer a cross-cutting perspective, but the mechanism of projecting the semantics into values to be displayed can serve as a powerful foundation for integrating a cross-cutting perspective with a local perspective. Similarly, the *Field* environment [3] offers general visualization of program behavior for live coding performances directly within the programming environment, thus potentially enabling a live cross-cutting perspective.

### 2.3 The Babylonian/S Programming Environment

We implemented our prototype as an extended version of the Babylonian/S programming environment [38], an example-based live programming environment in Squeak/Smalltalk. There are two core concepts relevant to the integration of the cross-cutting perspective: *examples* and *probes*.

Programmers can annotate Smalltalk methods with examples, which contain necessary objects to call a method (see Figure 4). These can either be method examples, for which programmers must provide scripts to initialize the receiver object and scripts to create the argument objects. Further, it can also be a generic script example, which can be any code snippet, that will invoke the annotated method. Programmers can name examples to communicate what they illustrate and configure set-up and tear-down scripts to handle any necessary system state.

To get feedback on dynamic behavior, programmers can then place probes on arbitrary expressions in the code [26] (see (B) in Figure 1). Probes record the results of every evaluation of the expression during the execution of an example. In the inline visualization of the probe, these snapshots are ordered according to their time of recording.

## 3 Integrating a Cross-Cutting Perspective Into a Live Programming Environment

To expand the experience of live programming to working on systems and debugging, we integrated call trace views as a cross-cutting perspective into the Babylonian/S live programming environment. We outline the features of the resulting environment and describe the underlying design decisions.

The major design challenge is the mismatch between the requirement of immediate feedback and the inherent complexity of call trace visualizations resulting from the complexity of the traces themselves. Depending on the task at hand, a different call trace tool offers suitable feedback [6]. Thus, we present programmers with three different kinds of views which are optimized for different usages. Further, the complexity of call traces also makes them difficult to navigate. Thus, our environment offers several means to navigate the cross-cutting perspective based on the local perspective. Also, we extended probes with path indicators and a visualization of control flow based on information from the call trace.

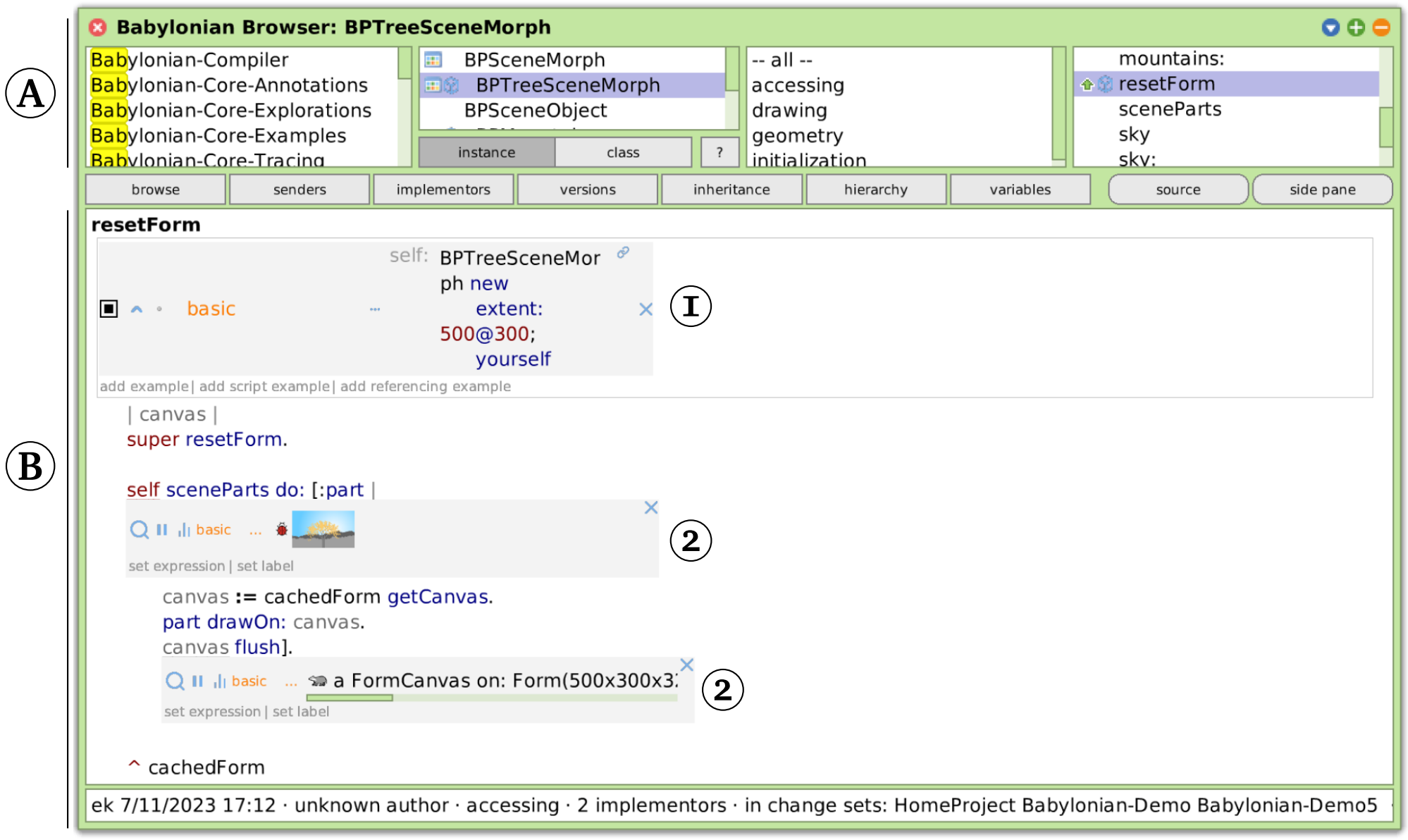

**Figure 4** The Babylonian/S code browser, based on the Squeak/Smalltalk code browser, without the new cross-cutting perspective tools. The standard Squeak/Smalltalk browser provides panels to select a method of interest (through selecting the package, class, method category, and lastly the method itself) (A) and a field for editing code, displaying the currently selected method (B). Included in this browser are core Babylonian/S features: The ability to define examples at the top of the code editing field (1) and two probes that display concrete information based on the example (2). In this specific case, a method example is used that has a small script to create an instance of the current class whose purpose is to draw a small scene with a tree and a background. The probes then display some of the variables used during drawing, in this case, one displays the canvas data visually and one displays a later canvas as data.

## 3.1 Three Views for the Cross-Cutting Perspective

Our environment provides three views of a call trace: full call tree, summarized call paths, and detailed call paths. All of them are based on a recorded trace of method invocations and probe hits during the execution of an example. They are live, meaning the displayed information is updated immediately on every change to the system. To mitigate cluttering with irrelevant information, the trace views can be filtered by programmers by selecting modules for which the method invocations should be included in the trace views. Our proposed views are general views of control flow, as the aim of this work is not to develop a new visualization of dynamic behavior but to integrate such visualizations into a live programming workflow. All three views can be accessed in the sidebar of the code browser. The top pane of the sidebar shows all active examples. The second pane shows the different views. For the summarized and the detailed paths views, a third pane appears, that lists all probes or methods invoked during the execution of the selected example.

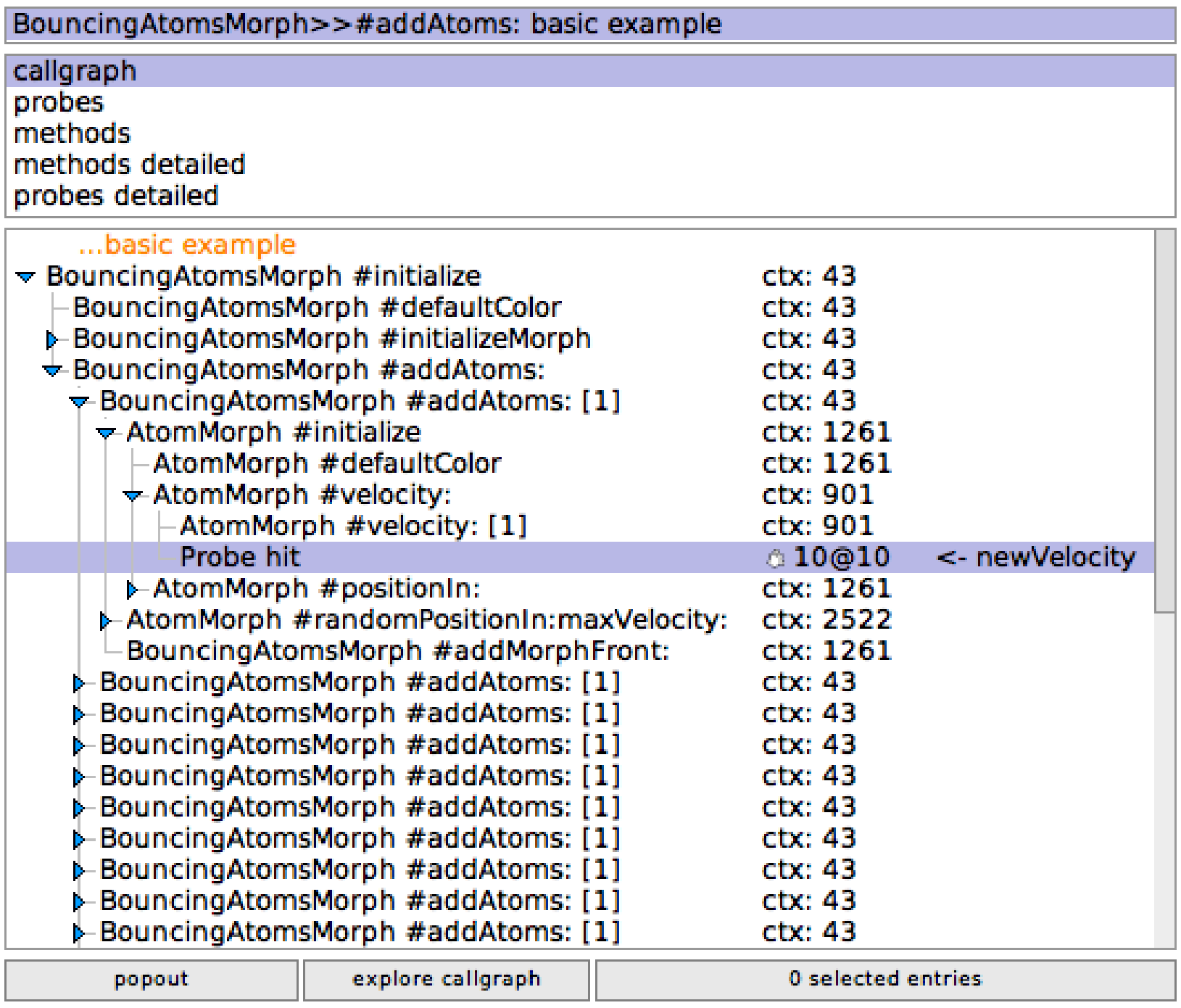

**Figure 5** The call tree view showing one entry per method, block, or probe invocation. The second column displays additional information such as the recorded probe value.

**Full Call Tree View** The *full call tree view* shows method calls and probe hits, nested according to caller relations (see Figure 5). The top entries in the tree are the calls occurring during the example execution, thereby this view can be interpreted as looking downward from the example into the tree. The displayed trace is exact in that it shows invocations of closures, including non-local invocations, and stack manipulations that can occur in Squeak/Smalltalk, such as resumed exceptions or generators.

For method calls, the tree also displays a number representing the identity of the stack frame, providing programmers with orientation in the case of co-routines. For probe hits, the tree shows the recorded values and the source code enclosed by the probe. To browse different invocations of a method, the context menu of tree entries allows one to jump to the next or the previous invocation of a method.

**Summarized Paths Views** Often, programmers might only be interested in the localization of a particular probe or method in the overall control flow. Therefore, we included the *summarized paths view* that can be interpreted as looking up from a single probe or method towards the example (see Figure 6a).

The summarized paths views display the different paths the control flow takes from the example to the probe or method. One summarized path represents a group of

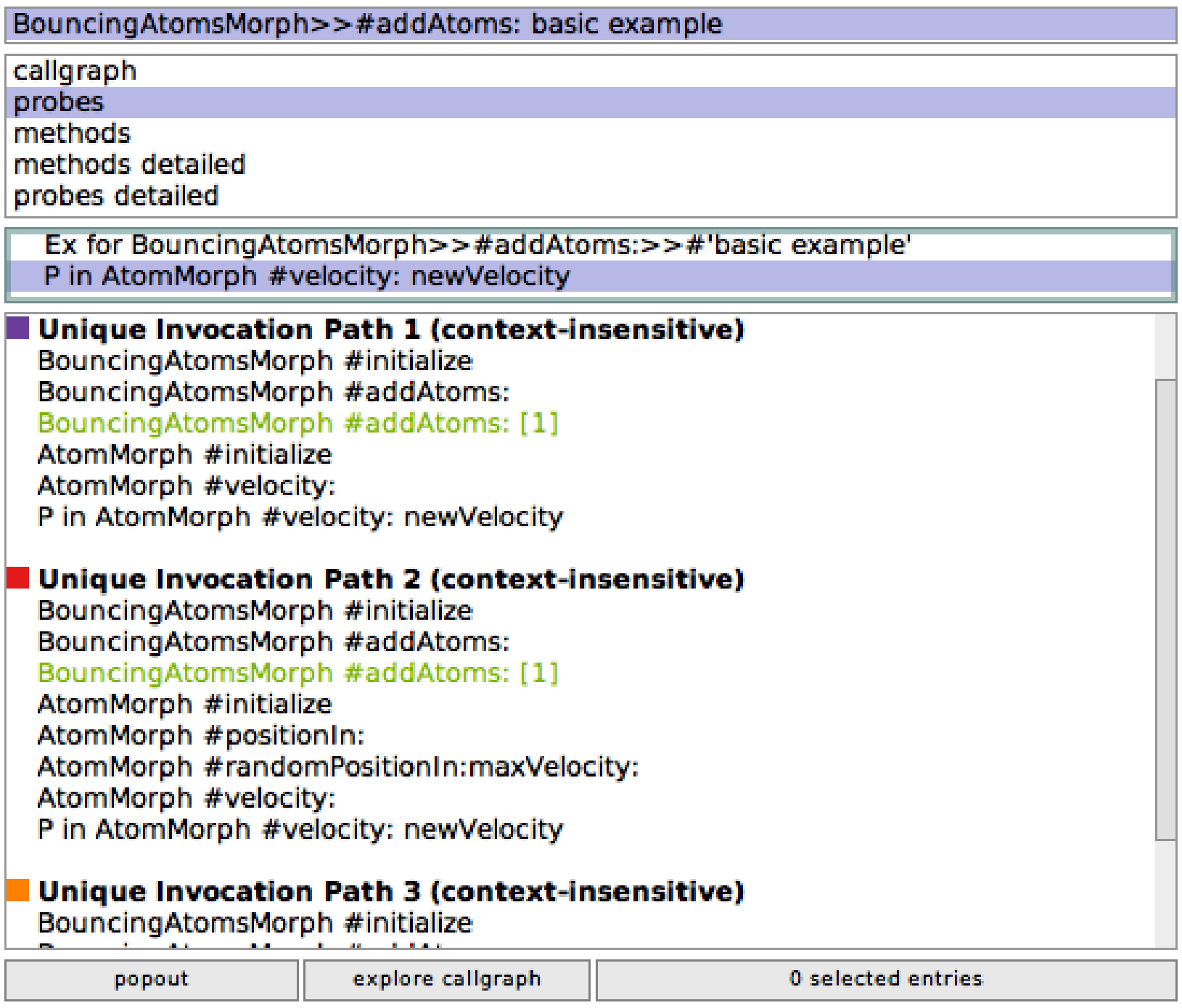

**(a)** The summarized paths view for probes. It shows the possible stacks whenever the selected probe is invoked. The sidebar also offers a corresponding view for methods.

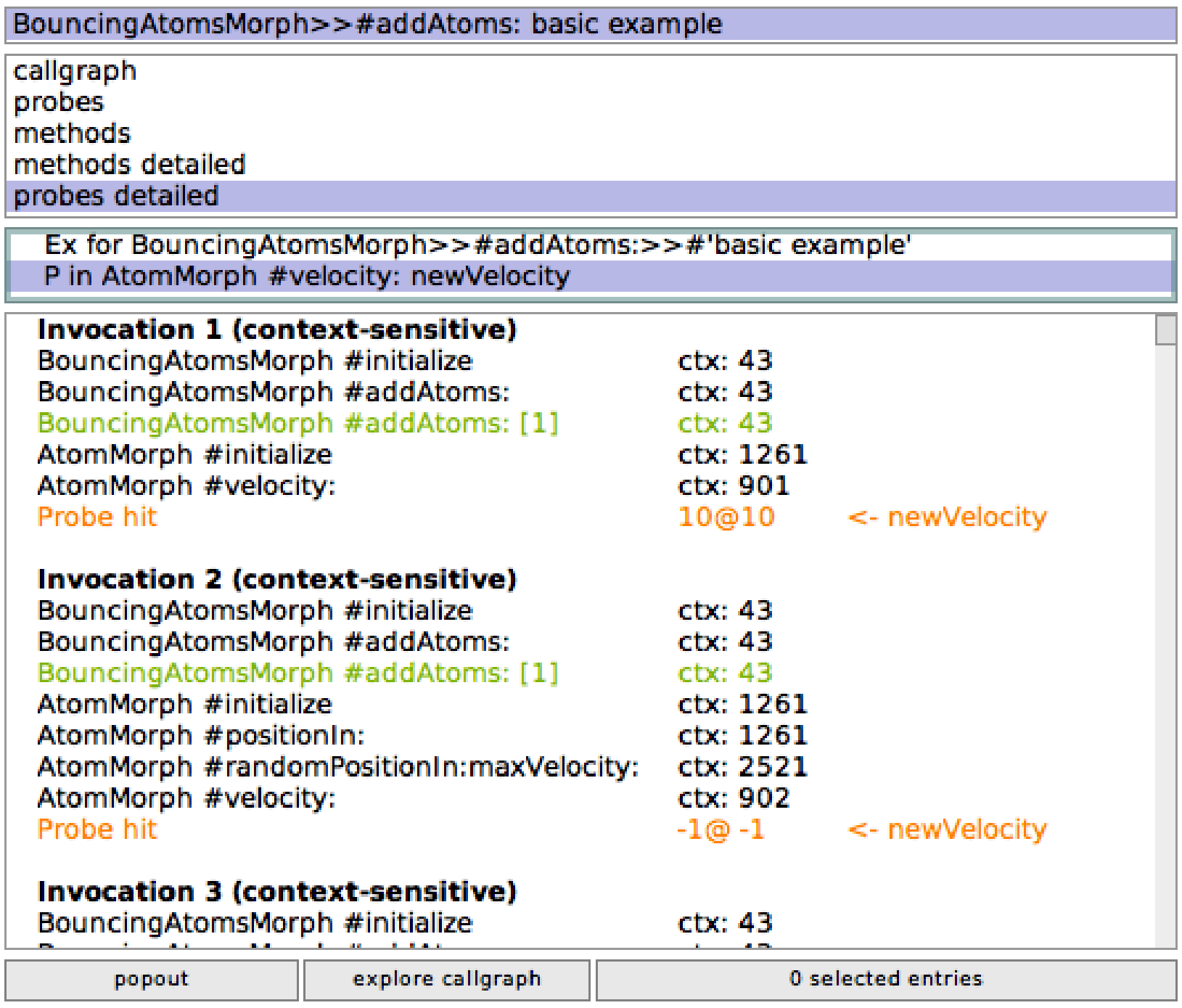

**(b)** The detailed paths view for probes. It shows all invocations of the probes and the stack at invocation time. The sidebar also offers a corresponding view for methods.

**Figure 6** The two paths views: summarized paths (a) and detailed paths (b).

invocations that all had a stack corresponding to the methods in the path [31]. Thus, for one path there might have been several invocations of a method or a probe that all had the same invocations on the stack.

To ease the comparison of paths, the common ancestor of all paths is highlighted in green. The section of the path above the common ancestor is the same for all paths.

**Detailed Paths Views** The summarized paths views show all actual control flow paths during an example. While programmers can get an overview with the summarized views, they can not display run-time data of individual traces along these paths. To see run-time data, programmers can select the *detailed paths view* for probes and methods (see Figure 6b). This view shows all traces leading to a probe or a method. In the case of the detailed probes view, each trace ends with a recorded value. Thus, this view can be used to inspect individual stacks for recorded values.

**Note on Response Times** As a short feedback loop is an elementary part of live programming environments, we briefly discuss the impact of the instrumentation on the system response time. The following is not a detailed performance analysis and should not replace one, but serves as a short characterization of the orders of magnitude. We use an instrumentation based on byte-code rewriting whose worst-case performance penalty when instrumenting the whole system including the standard library is a factor of 23 with an average performance penalty of about a factor of 10. This makes live feedback still feasible, as examples are small in scope. As long as they run for less than 100 ms, they will on average finish within 1 s when every single call in the example is recorded. Further, these numbers are upper bounds as they assume that the complete runtime system is instrumented. Focusing the instrumentation on methods outside of the standard library can reduce the number of calls to be recorded considerably and thus the run-time overhead.

### 3.2 Combining the Local and the Cross-Cutting Perspective

To make the information on the cross-cutting perspective directly available, we support jumping between the code and the trace views. Further, to provide programmers with initial information on the control flow without requiring them to switch to the trace views, our environment offers inline path indicators and inline visualization of value succession.

**Jumping Between Code and Trace Views** Programmers can navigate from elements in the source code editor to entries in the trace view and vice versa.

To contextualize a single probe value within the overall control flow, programmers can display the corresponding entry in the trace view via the context menu of the value (see (A) in Figure 7). This will select the corresponding probe invocation entry in the full call tree and the detailed paths view. Jumping to the summarized paths view will select the corresponding path. Further, programmers can jump to the call trace entry of the first invocation of a method through the context menu of a method. Navigating to other calls of the method can be done via the navigation actions described above.

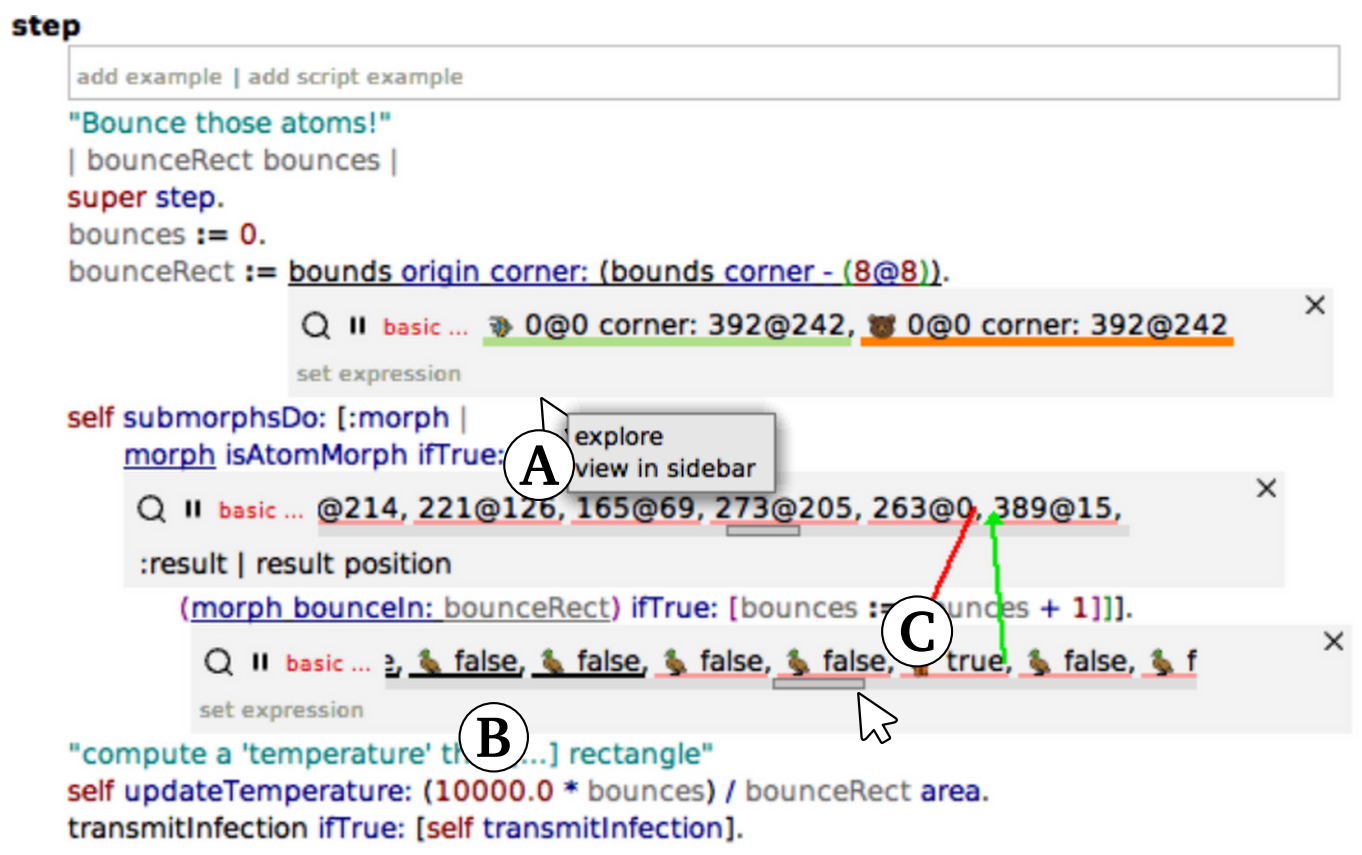

**Figure 7** The three combinations of the local and the cross-cutting perspective. Programmers can jump from probe values to the entries in the currently open trace view via the context menu of a probe value (A). The colored bars below probe values indicate the path that led to the probe invocation recording this value (B). When hovering over a probe value, the editor shows arrows to the probe values in the same method that were recorded before and after the hovered probe value (C).

Programmers can jump from every method entry in a call trace view to the method implementation by double-clicking on the entry. Doing so on a probe entry will open the method containing the probe. Further, programmers can directly inspect probe values by double-clicking on probe values displayed in the call trace views. This will open an interactive object explorer, which allows programmers to inspect the internal state and invoke methods on the object.

**Inline Value Path Indicators** To reduce the need to switch to the call trace views, probes indicate the summarized path that led to a recorded value as a colored bar below the value (see (B) in Figure 7). Thus, programmers can easily determine whether values result from different paths. The colors of the bars correspond to the path icon colors in the summarized paths view for probes (see Figure 6a).

**Inline Visualization of Value Succession** The full call tree view allows programmers to determine the temporal succession of any two probe values. However, this requires switching to the call tree view and manually navigating to the two probe values. To allow programmers to directly see the temporal relation between the probe values within one method without the need to switch views, we visualize this relation (see (C) in Figure 7). On hovering a value, two arrows are shown, indicating which value was recorded before the hovered value and which one afterward.

## 4 Walk-Throughs

One of the goals behind Babylonian/S is a better connection and visualization of the relationship between source code and its run-time behavior [38]. To tighten that connection, programmers define examples inside their source code that automatically run in the background whenever code is changed. Programmers usually have no insight into the behavior of the code covered by an example (except for possible side effects caused by the example execution). To gain insight into the run-time behavior of an example, Babylonian/S offers probes, assertions, and replacements, which we refer to as *annotations*. These annotations can give programmers insight into the example execution process at manually selected points during its execution.

As of now, programmers are forced to either step through the example in a debugger or increase the number of probes to get a more overarching picture of the code. For example, programmers who are interested in which procedures are called from a procedure could run the entire example in a debugger or place probes in a variety of possibly executed procedures.

When using multiple probes, it can also be difficult to understand the relationships between multiple probes.

Based on these observations, and our own experience with working with Babylonian/S, we determined the following use cases where Babylonian/S could benefit from call graph information (each use case is scoped to the execution of a single example):

1. Which procedures/annotations are executed?
2. Given a procedure/annotation, when is it executed during the example execution?
3. Given a procedure call, what other procedures are called from it?
4. How do all executions of a single procedure/annotation relate to each other (both temporal and structural)?
5. Given two or more procedure/annotation executions, how do the executions relate to each other?

We aim to aid users in these use cases by providing them with call graph exploration tools that further narrow the gap between source code and run-time behavior. We present two walk-throughs that show how our additional call-graph-based tooling can aid programmers in those use cases.

For all walk-throughs, call graph tracing was activated for (i) all methods located in `Morph`, `Model`, and their subclasses (ii) all methods in the `RealEstateAgent` and `Flaps` classes (iii) the methods that contain the examples themselves. We deliberately did not instrument all methods, so that the call graphs remain manageable in size and the call tracing has less of a performance impact. We envision that the partial instrumentation of only a subset of methods reflects how programmers will commonly interact with our tools.

**Walk-Through 1: Use Cases 1 and 4** A user working on the "bouncing atoms" simulation is interested in how and when the position of atoms is updated. Thus, they are interested in the relation between calls to the `Morph >> position:` method that are made as part of the "bouncing" example. They open the method in the Babylonian Browser (see Figure 8). No example is currently selected in the list of active examples in the sidebar (area A; this means that areas B, C, D, and E are currently all empty).

The user now right-clicks on the method name (step 1) and selects "show in sidebar" in the popup context menu (step 2). Since no example is currently selected, this opens

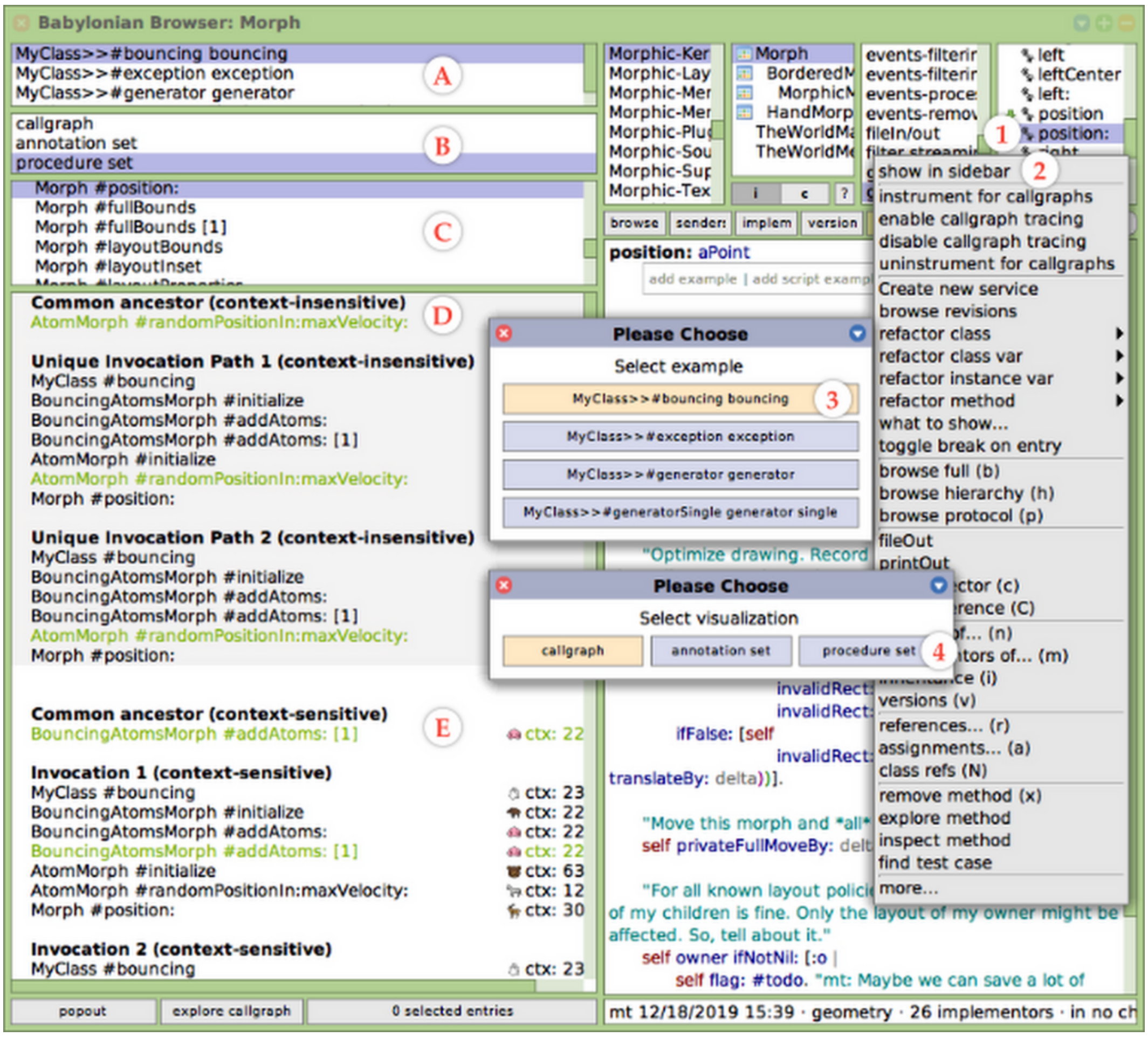

■ **Figure 8** Walk-through 1: How do all calls to Morph >> position: during the "bouncing" example relate to each other? Which procedures were executed during the example?

another popup that lets the user choose which example they are interested in. The user selects the "bouncing" example (step 3). Another popup opens that allows the user to select the desired visualization. They decide to open the "procedure set" visualization (step 4). Now, the sidebar is updated to highlight the selected example and visualization (areas A and B). Additionally, the call graph structure is generated from the last recorded call trace of the example, if it has not already been generated. Once finished, the visualization loads in the sidebar (areas C, D, E).

The selected procedure set visualization is split into three areas: Area C shows all procedures that were called during the example execution. The user-selected procedure, Morph >> position:, is pre-selected. Areas D and E show the context-insensitive and -sensitive invocation paths and ancestors of all calls to Morph >> position:.

By looking at the context-insensitive common ancestor (area D), the user learns that all of the calls to Morph >> position: during the example are coming from the method

AtomMorph >> randomPositionIn:maxVelocity: (4). By looking at the context-insensitive invocation paths, they learn that that method is called from AtomMorph >> initialize and BouncingAtomsMorph >> addAtoms: [1].

Looking at the context-sensitive ancestor and invocation paths (area E), the user sees that the context-sensitive ancestor of all calls to the Morph >> position: method is BouncingAtomsMorph >> addAtoms: [1]. This means that the execution of the addAtoms: method with context identifier 22 is responsible for all calls to Morph >> position:. This is expected since the example consists of just a single call to BouncingAtomsMorph >> new.[2]

By scrolling to the bottom of area E, the user can look at each individual context-sensitive invocation path and also see the total number of invocations of Morph >> position:.

Now that the user selected the example in the sidebar, they can also easily see the set of procedures (area C) that were executed during the example execution (1). The set of Babylonian annotations can be accessed by switching to the "annotation set" visualization (area B).

**Walk-Through 2: Use Cases 2, 3, and 5** A user is interested in how and where the 11 buttons of the Squeak/Smalltalk class browser are created.

The class browser uses the ToolBuilder framework, which allows programmers to describe the graphical user interface of a tool independent of a specific graphics framework. In the ToolBuilder framework, programmers first describe the general structure of the graphical interface of a tool through specification objects. After constructing the complete specification object graph, the specification is passed to a specific ToolBuilder class that creates the graphical objects for the corresponding graphics framework.

The user has found one method that constructs specification objects for the buttons in the class browser, but the method does not list all visible buttons. Also, with static navigation tools, they get too many candidate methods and turn to examples. They write an example that opens a class browser and immediately closes it so that running the example does not pollute their screen with an additional instance of the browser. The example can be seen in Figure 9. The button rows the user is interested in are labeled with A and B.

First, the user clicks on the "view in sidebar" button on the example (step 1), which opens a popup that asks the user which visualization they want to open. The user selects "callgraph" to open the call graph visualization (step 2). Because the example is fairly large, converting the call trace into the call graph takes a noticeable delay of about one second. The sidebar now shows the call graph of the example (area C). The tree widget initially displays the call graph fully collapsed, which means that just three root nodes are visible (the example node, the MyClass >> browser enter node, and its corresponding exit node).

[2] If the example instead contained multiple calls to the BouncingAtomsMorph >> new method, the context-sensitive closest common ancestor would be the example itself since the addAtoms: method would be invoked twice with different context identifiers.

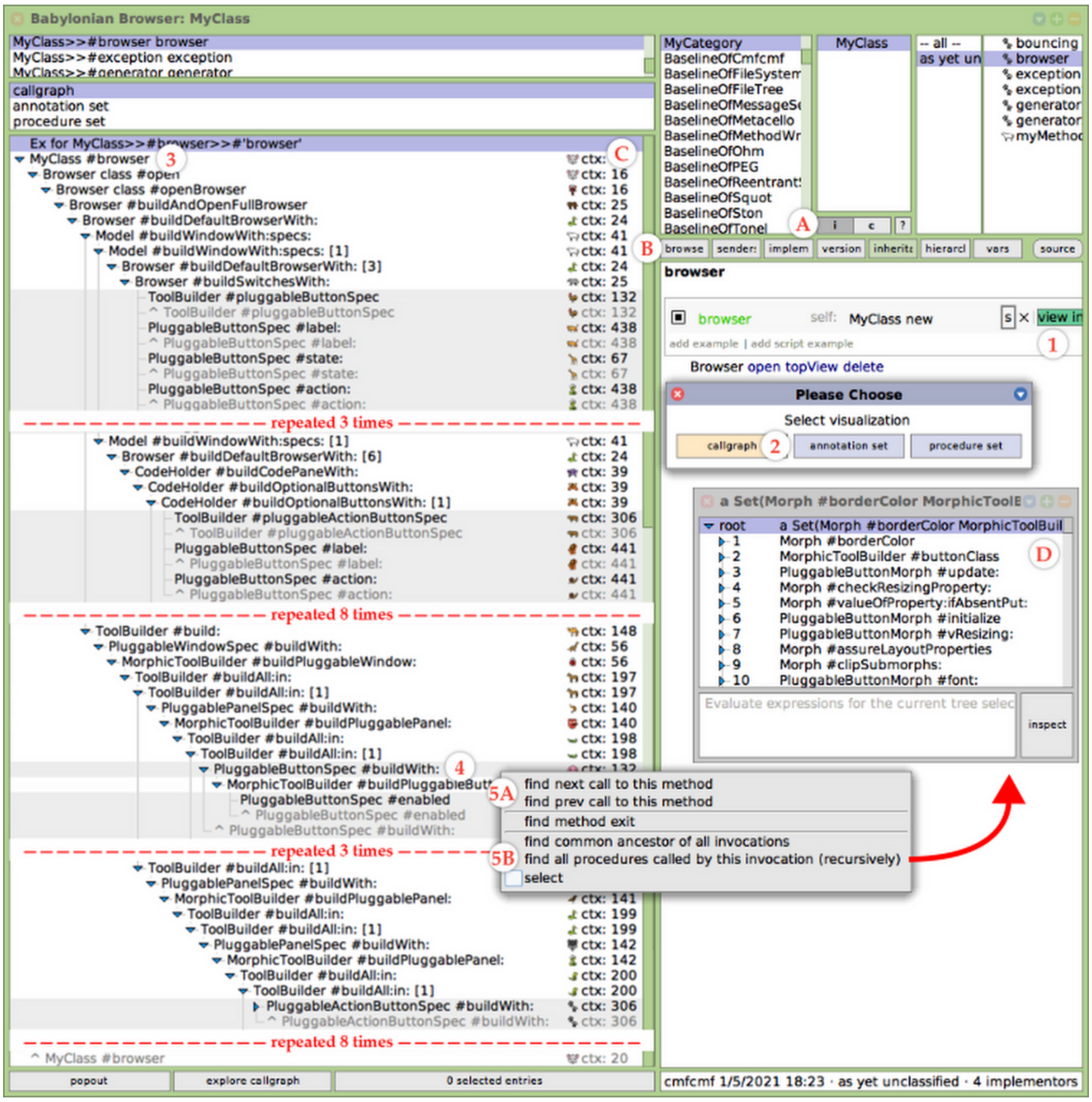

**Figure 9** Walk-through 2: Which methods related to *ButtonSpecs are called when opening a browser? What methods are called when building a button from a PluggableButtonSpec? To fit in this figure, we removed repeated rows from the call graph (indicated by the red “repeated n times” labels).

From their experience, the user knows that the class browser uses ToolBuilder and that buttons in ToolBuilder are created using Pluggable(Action)ButtonSpec classes. Therefore, they are interested in any calls that involve these classes (5). The user could manually expand the entire call graph recursively to look for any use of these classes by using the blue arrow icons at the start of each row.

However, a simpler alternative is to use the standard Squeak/Smalltalk tree filtering mechanism: When the call graph is in focus, users can apply a filter to the tree simply by typing. Knowing that, the user focuses the call graph and types "buttonspec" (step 3). This expands the entire call graph but makes all nodes invisible that do not match the filter and do not have any (recursive) children that match it. This greatly reduces the size of the call graph compared to an unfiltered graph. Matching nodes have a gray background color.

The user is now able to focus on the uses of the classes of interest and can determine that there are four phases in which ButtonSpec classes are used. First, three ButtonSpec objects are created (likely for the three buttons labeled as A), then, another eight ActionButtonSpec objects are created (likely for the eight buttons labeled as B). Later, the buttons are built from the ButtonSpec objects: First the three top buttons, then the eight bottom buttons.

If the user is interested in one specific method, like PluggableButtonSpec >> buildWith:, they can open its context menu using right-click (step 4). Using the "find next/prev call to this method" options (step 5A), the user can then jump around the call graph to explore all calls to the method (2).[3]

Knowing where the buttons are created, the user now wonders what procedures are involved in the creation of a button (3). To learn about that, they re-open the context menu (step 4) and use the "find all procedures called by this invocation (recursively)" option (step 5B). This opens the set of methods that are called by the selected method call in an explorer window (area D).

## 5 Exploratory User Study on the Effects of a Cross-Cutting Perspective in Live Programming

The introduction of a cross-cutting perspective is a significant change to a live programming environment. We conducted an exploratory, observational user study to examine how a cross-cutting perspective might affect the workflow of programmers. We decided to conduct an observational study instead of a controlled experiment to test the effects of the cross-cutting perspective, due to the lack of a solid theory that can be used to generate suitable hypotheses.

We expect the call tree view to be useful to programmers in particular when debugging unexpected behavior or when exploring an unknown program. Hence, to

[3] Alternatively, the user could also use the combination of class and method name to filter the entire call graph as they did previously to select all nodes that match it. Or, they could open the procedure set visualization to inspect all invocation paths of the selected method.

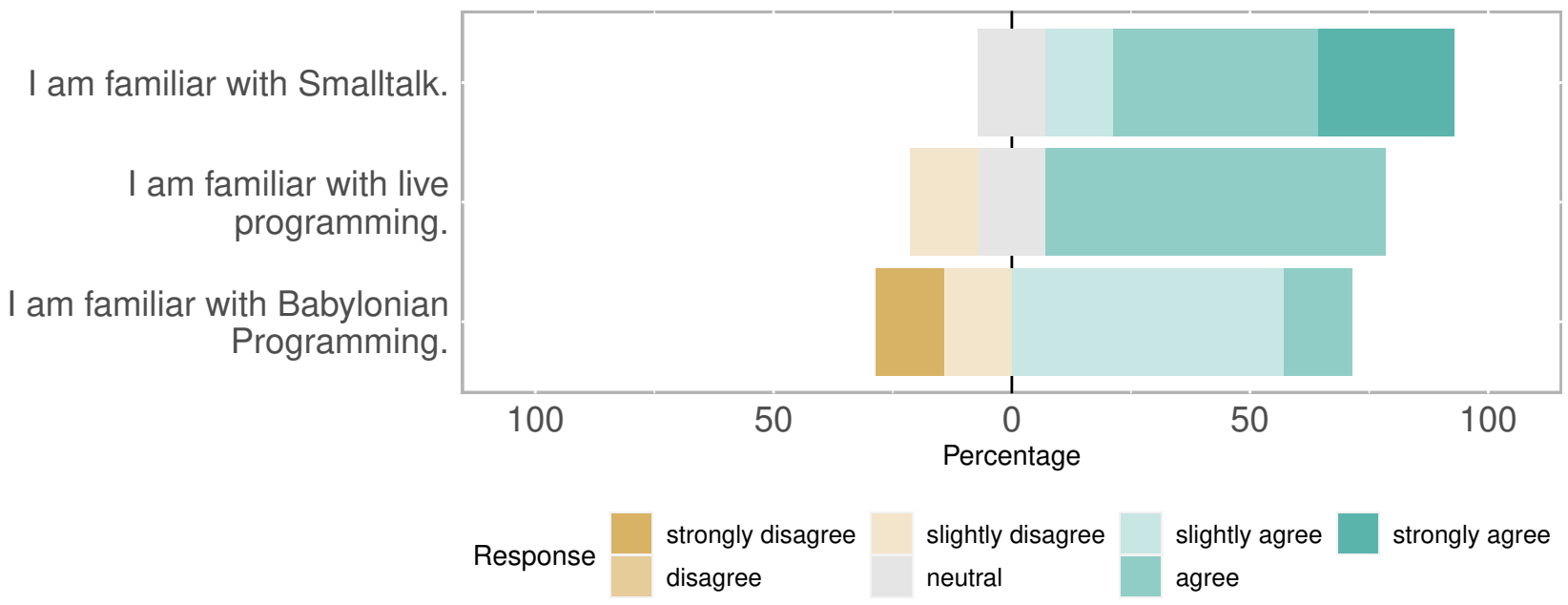

**Figure 10** Results of the background questionnaire shown as a stacked bar chart of the percentages of responses. Participants agreed that they were familiar with Smalltalk and live programming. Also, participants reported to be somewhat familiar with the Babylonian Programming environment, but not all of them.

provide study participants with opportunities to use the new tools in our cross-cutting perspective, we decided to focus on debugging tasks. In this scope, we explored the following research questions:

**RQ1:** Do programmers find the cross-cutting perspective helpful?

**RQ2:** For which activities do they use which perspective?

**RQ3:** How do programmers navigate the cross-cutting perspective?

### 5.1 Procedure

We investigated these questions in a think-aloud user study. We recruited 7 graduate-level university students from our faculty (5 male, 2 female). They reported 3.5 - 12 years of programming experience, including 0.5 - 4 years of professional programming experience. More specifically, most agreed that they were familiar with the Smalltalk programming language and live programming (see Figure 10). This is a result of the fact that the Squeak/Smalltalk live programming environment is used in the project work of two compulsory undergraduate courses. Thus, while they can be considered junior software developers with regard to their professional programming experience, they are experienced with live programming tools. Participants reported being somewhat familiar with the Babylonian Programming environment, presumably as they had seen a brief demonstration of it in a lecture (see Figure 10). However, none of them had used it themself before the study. Further, due to the demonstration, the participants might have associated the Babylonian Programming environment with the examiner. We aimed to prevent Hawthorne effects, resulting in participants adjusting their answers due to sympathy towards the examiner. Therefore, we told participants that the cross-cutting perspective was developed by a non-disclosed student as part of their completed master's thesis and we were considering whether to pursue the work further.

Sessions of our study were conducted online via Zoom. Participants used their computers running a pre-configured Squeak/Smalltalk environment, ensuring that all participants employed the same programming environment. Each session took 2.5 hours and started with a 20 minutes hands-on tutorial on the features of our environment followed by a short explanation of the think-aloud method. The participants then started to work on three tasks. All participants worked on the tasks in the same order. After completion of one task, we asked participants to continue with the next task. Each task was introduced by first demonstrating the failure, then explaining the application domain, and finally pointing the participant to the initial example.

After 90 minutes of working on tasks, we started the debriefing regardless of how many tasks they completed. In the debriefing, we asked them to fill out a questionnaire about the cross-cutting perspective in Babylonian/S. The questionnaire is based on the utility section of the USE questionnaire [8]. After that, we conducted a semi-structured debriefing interview. The interview includes two questions for which we asked participants to arrange features written on virtual cards according to their utility. The first of these questions comprised the integration of the local and the cross-cutting perspective, and the second listed the three different kinds of cross-cutting perspectives as well as probes in general as a representation of the local perspective.

**Detailed Description of Tasks** Participants worked on three tasks.[4] All tasks were in domains new to participants. As starting points, we created an initial example for each task. That example executed a script including steps that we used to manually demonstrate how the program to work on failed. Further, the tracing was scoped to the packages of the classes that were used in the initial example. We designed the tasks so that each required a different debugging approach.

The first task is to repair a defect in a small graphical gas tank simulation. To find the defect, programmers have to first identify an argument as erroneous. Then they have to trace its origin up the stack and down the stack again to find the erroneous expression. As a starting point, we provided an example that executes one step in the simulation. The minimal number of classes that the participants need to understand for this task is 2, which together contain 41 methods with a total of 482 lines of code.

The second task is to find a defect in an interactive programming tool that displays dependency information for packages. Repairing this defect requires programmers to find the initialization of the erroneous state. This erroneous state does not directly lead to the observed failure but only later after a user event. Thus, directly examining the stack at the time of failure will not suffice. As a starting point, we provided an example that mimics the user interactions of selecting the package and the class that led to the failure. The minimal number of classes that the participants need to understand for this task is 1, which contains 50 methods with a total of 758 lines of code.

The third task involves debugging an indirect recursive evaluation of a method. This method is a rule of a visitor working on an abstract syntax tree of a simple expression language that can be styled with additional markup. Debugging this involves finding the method that creates the styling objects in the first place and understanding that the problematic method is recursive. As a starting point, we provided an example that includes the parsing of a simple expression in the expression language and the application of the styling visitor. The minimal number of classes that the participants need to understand for this task is 1, which contains 17 methods with a total of 133 lines of code.

[4] We have published the task materials at https://doi.org/10.5281/zenodo.10461488 (accessed February 20, 2024).

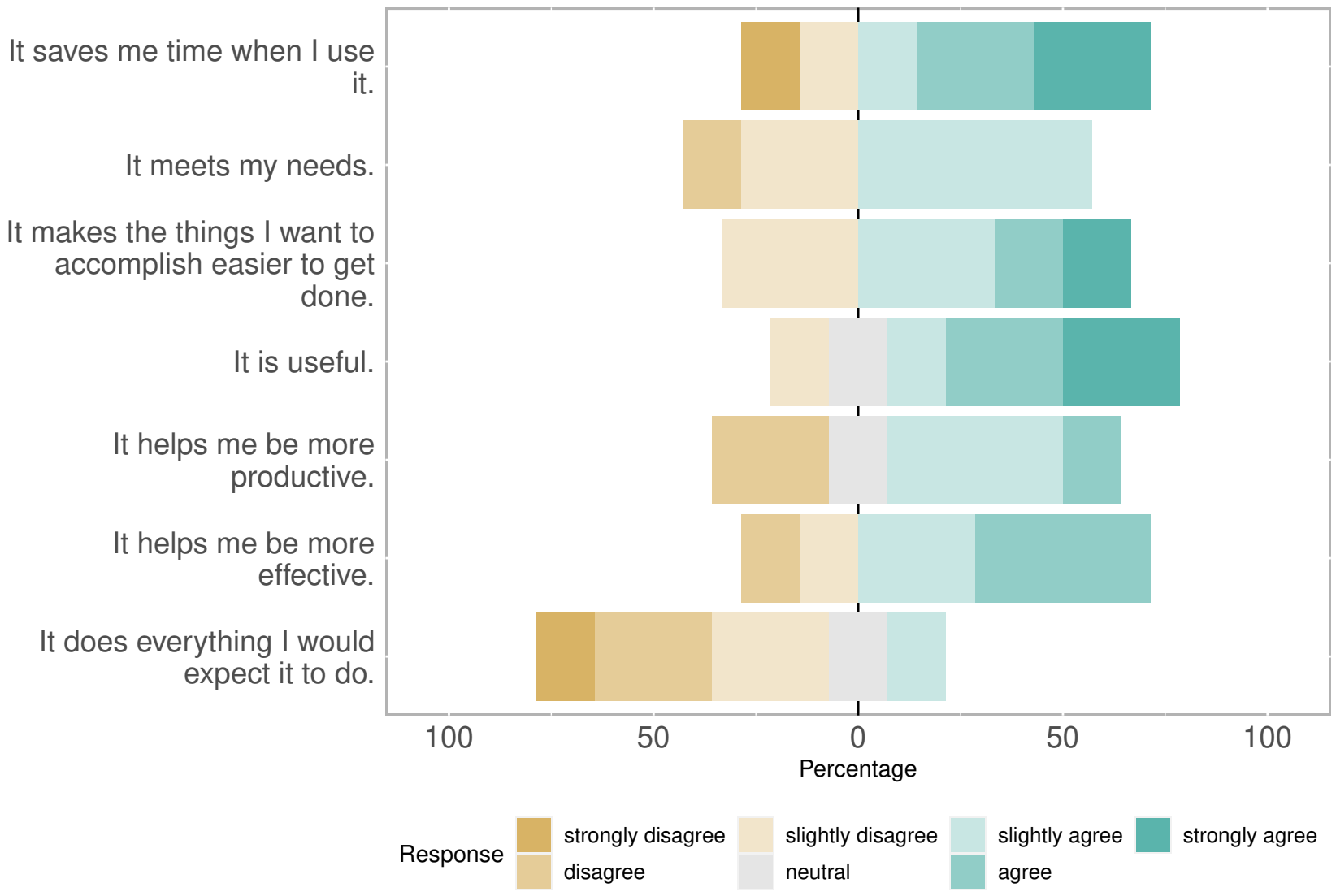

**Figure 11** Results of the utility questionnaire shown as a stacked bar chart of the percentages of responses. The questions refer to the cross-cutting perspective and its integration into the local perspective. Most participants agreed that it was helpful and that it makes things easier and saves time. At the same time, they reported that the tool did not provide everything they expected.

### 5.2 Utility of Cross-Cutting Perspective (RQ1)

Results of the utility questionnaire show that the majority of the participants perceived the live programming environment in combination with the cross-cutting perspective as generally useful (see Figure 11). In the debriefing interview, participants explained that they found it most useful for debugging and navigation (for details see Section 5.3). With regard to the cross-cutting perspective itself, they mentioned typical advantages of trace-based tools such as seeing all of the invocations at once or being able to go back in time. Many participants felt that the tool did not yet do everything they needed it to (see last question in Figure 11). We discuss the missing features throughout the following sections.

**Cross-Cutting Perspective as Extension** With regard to the integration of a cross-cutting perspective into a live programming environment supporting probes, many participants treated the cross-cutting perspective as an add-on to the probes and the probes as the primary feature of the environment. This is for example reflected in the answers to the utility-rating questions, for which all participants rated the probes as the most useful feature. P1 also articulated this explicitly in their interview: *"If I had not had the probes, the call tree view would have been not as helpful."*[5]

**Alternative Tools** To determine which tools our environment subsumes, we asked participants which other tools they would have used to solve the tasks. Most of them mentioned the step-wise debugger, static navigation tools for browsing senders and implementers of methods, and the Squeak/Smalltalk object inspector. Thus, for our participants our proposed environment seems to offer features similar to what common debugging and navigation tools provide.

**Participants with Negative Responses** P1 and P2 disagreed with all utility questions in the questionnaire (see Figure 11). During the interview, we determined that P2 had not yet used any trace-based tool during programming beyond printf-debugging. As a consequence, they struggled to make sense of the presented information. Further, P2 expected the tool to display information that is more immediately useful to them: *"I expected to be able to see more in it than is actually possible. I tried to see something directly in it. That was not very successful."* P3 was overwhelmed by the amount of information presented and as a result, they resorted to their usual workflow of using static navigation tools and a step-wise debugger: *"This [using the debugger] probably also happened as there is so much text and there are so many widgets on the right that I felt a bit overwhelmed by them."*

**Little Anticipated Benefit for Writing Code** Finally, we assumed debugging to be the use case in which the tool is most useful. Ideally, they would also be able to use the cross-cutting perspective to spot misunderstandings early on during writing their program. However, as the views still contain many elements and thus require programmers to focus on them to interpret them, we do not expect the cross-cutting perspective to yield much benefit for writing code yet. To test this assumption, we asked the participants whether they would use the cross-cutting perspective while writing code.

All participants answered that they would primarily use the probes for writing code. Some said they might occasionally look at the cross-cutting perspectives when working with complex control flow or when extending existing code. For example, P4 said: *"I think the tools are more suitable for getting insights about the code […] It might be helpful if you're hooking into something, if you're extending code, or if you're hooking into a module somewhere. But if you really write completely new code, then you don't need much code understanding."* At the same time, most participants stated they could not imagine using the cross-cutting perspective for writing code but would make use of it when unexpected program behavior occurs. P3 explained it as *"I don't think [that I would use it to write code] because when I'm writing code myself I have this internal model in my head […] When you're designing behavior yourself, you're so much into the subject that you wouldn't use the right [the cross-cutting perspective]. I would really only*

[5] All quotes have been translated from the participants' native language to English.

*use the right [the cross-cutting perspective] when I notice that it is not doing what it's supposed to do and then to find a fault."*

These statements confirm our own assessment, that the cross-cutting perspective is primarily used to understand behavior in retrospect. A potential avenue for future work to improve the usefulness of the cross-cutting perspective for writing code may be to devise a compact graphical visualization that is easier to consume.

### 5.3 The Two Perspectives for Debugging and Navigating (RQ2)

With regard to activities in the different perspectives, the participants who found the environment useful mostly used it for the common debugging workflow of identifying erroneous state and tracing it from the failure of the program to the defect in the source code [46]. In general, it is not surprising that participants employ dedicated debugging techniques, given that they worked on debugging tasks. However, more specifically, we could also observe that participants employed the cross-cutting perspective as a focused navigation tool.

**The Two Perspectives During Debugging** We observed that all participants used probes to find erroneous state as a first step during a task. Among others, P2 explicitly mentioned this during the interview: *"This [finding faulty state] is what I use debugging tools for, so going in and looking at all the state first. Most of the time that's a good technique, and of course, probes are handy for that."* Once they deemed some state as suspicious, they tried to find the origin of the erroneous state by navigating the trace with a combination of a call trace view and probes. When they found a suspicious method, many started to slowly read its source code and occasionally checked hypotheses by placing probes. P1 described this overall workflow as *"I noticed that it is very similar to the standard workflow with the debugger and the object inspector, but it has the advantage that I can look at all call stacks at the same time."* While almost all participants likened their workflow in our environment to their workflow when using a debugger, some participants noticed the benefit of not having to restart and reset their debugging tools. For instance, P6 stated: *"I used the tool to see how the call came about. Typically I would have to open a debugger and step to it. The [tool] took that step off me."*

**Cross-Cutting Perspective for Navigating** Within this general approach, participants used the cross-cutting perspective primarily for navigating between methods relevant for the task at hand. Many stated that they would have normally used the senders and implementers tools for the same task, but that the cross-cutting perspective was more helpful as it only displayed the methods that were actually executed. This is particularly relevant as Smalltalk employs dynamic dispatch and navigating several levels of polymorphic calls usually requires programmers to keep track of the involved classes. P5 explained it as *"Probes show values and the sidebar shows control flow, which method is called from where and how. I could otherwise only find out about that by looking at what the senders are. [...] But those wouldn't be the actual senders in the* *example. It is very good, it shows me information that would otherwise require effort to get."*

**Preferred Views** During all of the tasks, we observed that participants used the full call tree and the summarized paths view for probes. Besides tracing the origins of state back in time, some participants also employed the full call tree view at the beginning of a task in a top-down manner to get an overview of the involved methods. Participants who employed the summarized probe paths view used it to determine which paths might be relevant. They did so by comparing differences between the paths. One participant went through the list of paths one-by-one, to exclude the ones that did not lead to the erroneous state. Very seldom participants opened the summarized method paths view and none used it a second time. One participant stated, that they felt it was redundant with the summarized probe paths. For them, the probes view served the same need as the methods view, as they always had probes in the methods they were interested in. This corresponds to our initial assumption that probes denote points of interest in a program execution.

None of the participants used the detailed paths views. This might be the result of how we designed the study tasks, but may also hint at a more general insight on the tool design. When asking questions on run-time values, participants were interested in specific values, shown in the probes. When asking questions about the control flow, they were only interested in the general paths, not specific ones. Thus, they used they preferred the summarized paths views over the detailed ones.

At the same time, one participant asked for a view displaying the values of all probes interleaved and ordered according to their recording time. They wanted to use this view to determine when inconsistencies in the run-time state first occur and how they spread. Several participants asked for some form of state-based cross-cutting perspective. They were mostly interested in seeing all reads and writes to instance variables and how they related to the overall control flow.

### 5.4 Navigating the Local and the Cross-Cutting Perspective (RQ3)

When working with the full call tree view, participants mostly navigate it starting from individual methods and probes using the features to jump from a probe value or a method to an entry in the call tree. P4 described this whole process as *"Jumping back and forth is a very good interaction. It feels super comfortable as a workflow […] You jump from the method to its entry and from the entry to an entry of another method and from the entry of the other methods you can jump back [into the source code]."* P7 characterized the experience as *"zooming out"* from a specific location and getting a broader view of the behavior in a call trace view. Correspondingly, after examining the call trace view, participants jumped back to a method they were interested in, which P7 characterized as *"zooming in"*. A notable exception to this were the occasions when participants explored the full call tree view in a top-down way at the beginning of a task, to get an overview notion of the methods involved in the trace.

For the probes view, we observed that participants did not use probes or methods as entry points, but instead used the paths in the probes view as a list of methods to be explored.

**Limited Usage of Value Path Indicators** Some participants explicitly mentioned that they used the path indicators below probe values as hints to potentially interesting control flow paths. However, none used the value path indicators in situations in which it directly displayed information relevant to a question participants were currently investigating. Even further, some participants were confused by them. Two participants assumed the colors to correspond in some way to the objects the methods are executed on. One of them clarified in the interview, that they were tracing the control flow "along" one instance, and while they knew that the colored bar did not encode the receiver object, the mismatch still confused them when trying to interpret the bars. This hints that for an object-oriented environment, the path indicators may become more useful when they would encode the object for which the method is executed.

**Limited Usage of Value Succession Visualization** The visualization of the temporal relation between probe values within one method was not used very often. Participants also did not use it in cases in which it would have provided information directly useful for the current question. For example, one participant was already investigating the method containing the fault with two probes, one showing objects from a collection and the other showing the results of the faulty transformation for each object. In this situation, the visualization of the temporal relation would have directly visualized that some transformation results did not match the original objects. Instead, they examined all probe values and checked whether they were consistent in relation to other values of the same probe. An interleaved view of probe values as suggested by one participant, might be beneficial for such situations.

## 6 Discussion

To put the conclusions from our study into perspective, we briefly examine the limitations of the study. Based on the prototype and the observations in our study, we discuss how live programming environments extended with a cross-cutting perspective can support programmers during debugging tasks. Finally, we discuss how live programming environments may in general serve as the basis for other, advanced tools for working with information on run-time behavior.

### 6.1 Limitations of the Study

We would like to emphasize that our study was exploratory and thus the presented results do not represent general observations about the effects of a cross-cutting perspective in live programming environments but as hints that may guide future designs of cross-cutting perspectives. In particular, the mere transfer of debugging techniques and the perceived utility does not imply that our environment will improve the workflow of programmers in any regard in comparison to other trace-based tools or a step-wise debugger.

Further, the characteristics of the participants might have limited the range of techniques we were able to observe.

First, while our participants widely differ in their self-reported years of experience, they are at the same time a homogeneous group with very similar traits with regard to their age group and their educational background.

Second and more specifically, the participants were already familiar with live programming in an exploratory-style live programming environment such as Squeak/Smalltalk in which all run-time state is accessible and explorable (see Figure 10). On the one hand, this means that the participants might be experienced with working with run-time data. On the other hand, this also might have kept them from applying new techniques, as the difference to their usual live programming workflow might have been small.

This also applies to the Babylonian Programming environment in particular. The students have previously observed a demonstration of the environment which did not yet include a cross-cutting perspective. This previous exposure to the environment might have shaped their experience and workflow.

### 6.2 Insights on Debugging in Live Programming Environments

Despite its limitations, our study also is an account of how programmers employ live programming environments for debugging. The fact that most participants could directly transfer debugging techniques to our environment is promising, as it hints that a live programming environment extended with a cross-cutting perspective can indeed replace the need for a separate debugger. At the same time, this is not too surprising, as live programming environments already work with information on run-time behavior. In particular, the combination of probes for identifying erroneous states and call trace views to trace their origin corresponds closely to common debugging techniques.

The resulting experience is, however, more similar to a back-in-time debugger than to that of a step-wise debugger [36]. While this is often beneficial, sometimes the in-time experience provided by step-wise debuggers might help programmers reason about a program more easily, as they follow the program step-by-step. How such an experience can be reproduced within a live programming programming environment remains a challenge for future studies and environments.

### 6.3 Adapting Live Cross-Cutting Perspectives to Other Programming Systems

We argue that the features and mechanisms we proposed are useful in the context of other programming systems. The reason is that the problem that motivated this work also occurs in all other programming systems that have any form of reusable behavior, whether it is procedures, functions, methods, or even macros. Programmers may reuse such behavior in different contexts. So, when programmers use the local perspective to investigate this reusable behavior, they may observe values that stem from different usages of the reusable part.

While the problem occurs in many programming systems, we also argue that features we proposed in this article can also be adapted to other programming systems. First of all, the basic Babylonian Programming concepts of examples and probes have been shown to work across several imperative languages, including Ruby and JavaScript [28] and different programming environments, namely Lively4 [38] and Visual Studio Code [28]. Further, the cross-cutting views only require basic tracing facilities to record dynamic information, and a notion of evaluation order that can be used to generate a tree of evaluations. Implementing cross-cutting perspectives in other programming systems would only require basic features for tracing and user interface extensions.

To gather the information necessary for our proposed views, programmers would need a mechanism to trace evaluations, as is available for most programming systems. The tracing needs to be complete, in order to show programmers the actual behavior and fast in order to support a live programming experience. The combination of complete and faster tracing can be difficult to achieve in some environments, but thanks to the small scope of example executions the overhead can be higher than what is acceptable for whole program executions. Finally, to support summarized paths views, programmers would need to define an equivalence relation that matches the evaluation mechanism of the language.

Integrating the cross-cutting perspective properly into the user interface of programming systems can be more challenging than implementing the tracing [28]. In many environments the integration of the views themselves may be simple, as many environments have side panes that display structural information. As an overview and navigation mechanism, the cross-cutting views we proposed fit well into these side panes. A major challenge is the integration between the local and the cross-cutting perspective. While the local perspective can be implemented using automatically added text or text augmentations that contain only text, the interactions between the local and the cross-cutting perspective, require support for arbitrary user interface widgets embedded into code, which is often not available.

Finally, for languages in which programmers do not directly express control flow, the cross-cutting perspective needs to offer different views. For instance, in logic languages such as Prolog, visualizations of the resolution may serve as a cross-cutting perspective [30].

### 6.4 Future Cross-Cutting Views in Live Programming Environments

Our observations have shown that programmers can already benefit from the proposed views that offer a cross-cutting perspective on program behavior. At the same time, all proposed views lack the directness of the dynamic information displayed in the probes. Instead, programmers have to consciously turn toward the cross-cutting views and work with them. Given that there seems to be value in a cross-cutting perspective, future views may make it even more useful.

Several new views may benefit programmers, based on the results of our exploratory user study. While no participant used the detailed paths views, one participant asked for a view showing the values of all probes interleaved and ordered according to their time of recording. They expected to use this to spot when the intermediate run-time state becomes inconsistent. This is similar to the timeline view proposed in DejaVu [15] and we have subsequently implemented a similar view [18]. The ordering along absolute time may also be visualized right next to the recorded values within the probe. This may already solve the problem of relating probe values causally, given that the number of values is low.

In addition to this interleaved view of probe values, programmers might have benefited from a view showing the changes to run-time state over time, similar to what OmniCode offers [13]. This might also be integrated with the full call tree view to show all locations in the trace when state was accessed. Similarly, to answer questions about state changes or to trace state infections during debugging, programmers might benefit from data flow views. As tracing state infections is a major activity of debugging, fine-grained dynamic slicing [32, 41] may allow programmers to navigate even more purposefully using the cross-cutting perspective [16].

An open problem remains the general complexity of views with a cross-cutting perspective on the program behavior which impedes immediate access to information, as programmers first have to orient themselves in the large number of visible elements. With the summarized paths views we aimed to reduce the number of elements that programmers are confronted with. Further summarization might make the information available more immediately. For instance, the visualization and querying ideas of multiverse debugging may be applied to multiple example executions, to give programmers a quick overview of essential methods for some behavior [25].

Finally, in object-oriented environments such as Squeak/Smalltalk, the views of the cross-cutting perspective and the indicators in the local perspective may be extended to also incorporate the current receiver object.

Studying the effectiveness of these new views, in particular whether they allow programmers to quickly explore and grasp cross-cutting behavior, remains interesting future work.

A general challenge for all future views remains the overhead introduced by the tracing infrastructure. While the small scope of example executions keeps executions generally short, the tracing infrastructure can slow down the execution so much that it does not support immediate access to run-time information anymore, and thereby breaks the experience of liveness. This particularly affects views that require the example execution to finish before they can display useful information. Possible solutions are streaming the parts of the view for which data is already available [4, 33], incremental computation that can speed up subsequent updates on changes to the code [26, 35], or just-in-time optimization of tracing statements [44].

### 6.5 Future of Live Programming: Foothold for Immediate Access for Advanced Dynamic Tools

Another way of looking at our environment is that it integrates a call trace exploration tool into a source code editor via the local perspective of a live programming environment. This illustrates how examples and the local perspective of live programming environments might serve as the foothold for other advanced dynamic tools. An example of an existing project that makes use of this potential synergy is the integration of a code synthesis tool into a code editor via the local perspective supported by projection boxes [7]. Based on our observations on how programmers debug in a live environment, we see the Whyline [16] debugging tool as a promising candidate to be integrated into a live programming environment.

When properly integrated, live programming environments such as the Babylonian Programming system, can shorten the feedback loop of those other tools, through short example executions and fine-grained feedback close to the source code.

Through the local perspective of such live programming environments, the feedback is available in the source code editor, which is the dominant view programmers use. In contrast, advanced dynamic tools, such as step-wise debuggers, back-in-time debuggers, or call tracers, are often implemented as separate perspectives that require context switches and thereby may hinder adoption. In contrast, the feedback generated by tools working with static information on programs is typically displayed directly within source code editors, for instance, syntax or typing errors displayed through code annotations

Further, live programming tools with examples can make feedback available faster by reducing the overhead resulting from instrumenting a whole program. For example, in the user study investigating the usage of Theseus [23], it was found that programmers did not use the tool often, as they avoided the complex setup: “Their feedback suggested that the overhead of instrumenting their entire project was too great, which may have been what kept [the tool] in the as-needed category of tool.” Similar results were found in a study on the usage of dynamic analysis at Siemens [45]. Explicit, user-provided examples could serve as generic starting points for tools working with dynamic information. As the examples represent a smaller scope to which the instrumentation needs to be applied, the dynamic analysis of the tools may run faster and consequently shorten the feedback loop.

We believe that a tighter integration of tools working with dynamic behavior into the live programming workflow, as we proposed in this paper, allows more programmers to benefit from them.

## 7 Related Work: Time-Travel Debuggers

Programmers can find defects in source code quicker by reasoning backward from a failure to the defect via the state infection chain, which is related to our cross-cutting perspective tooling. Time-travel debuggers support this strategy by allowing programmers to step the execution backward in time [22, 33, 34]. Building upon this elementary time-travel feature, they often provide additional advanced debugging features [33, 36]. Some of these advanced features also put programmers above time and allow them to see the overall behavior of the system. For example, the Path Finder debugger provides an integrated visualization of the call trace and the source code of the executed methods [33]. Nevertheless, the main perspective of time-travel debuggers often remains the in-time perspective of following along the execution

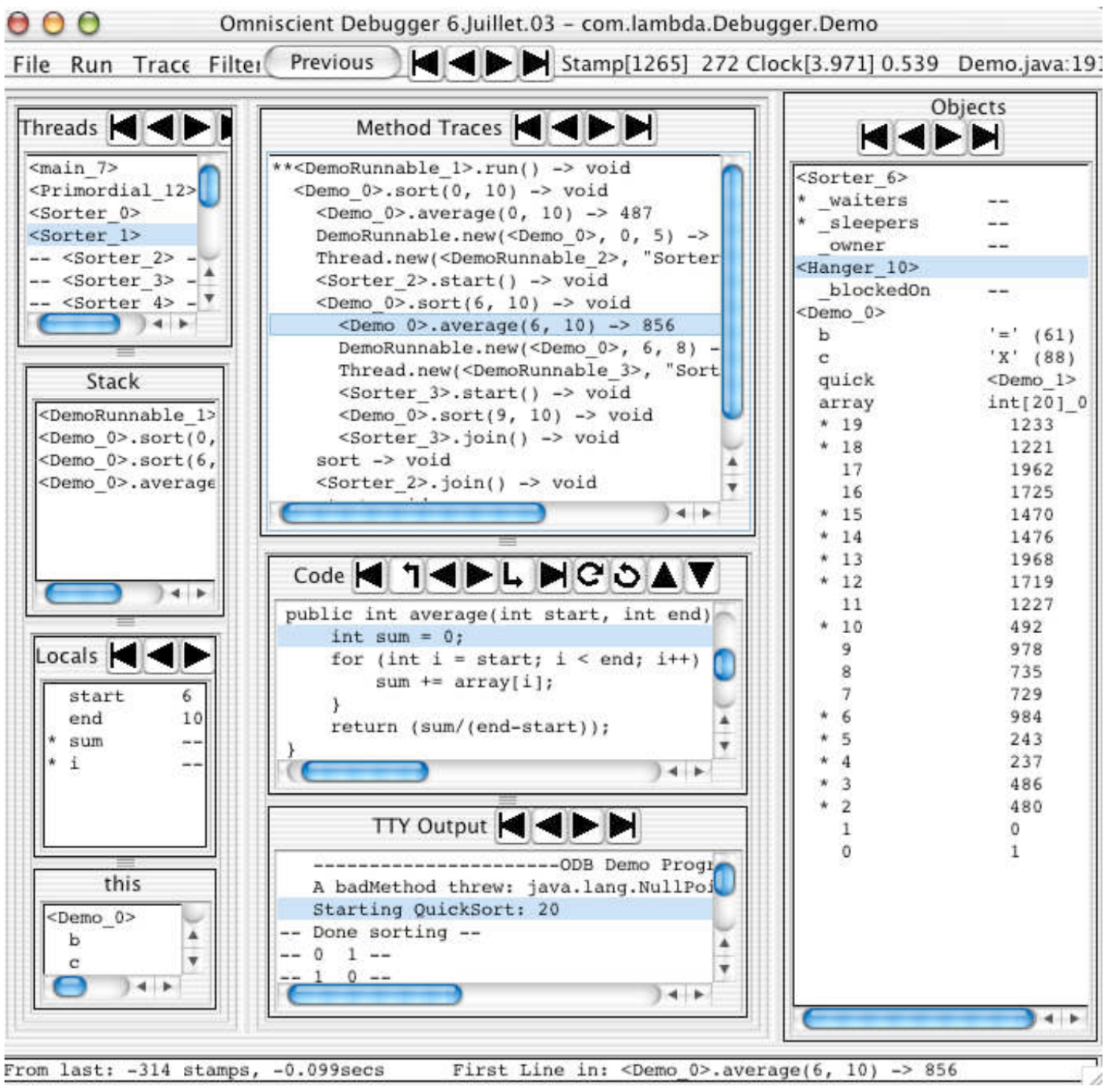

■ **Figure 12** A screenshot of the time-travel debugger called *Omniscient Debugger* [22]. The debuggers show a trace of the method executions ("Method Trace"), the code of the currently selected method execution ("Code"), the current state of local variables ("Locals"), the state of the receiver object ("this"), the stack at the point of execution ("Stack"), the current collection of active threads ("Threads"), the output of the program ("TTY Output"), and a state inspector for user-selected objects ("Objects"). All panes with arrows at the top can be used to navigate through execution time.

step-by-step, either forwards or backwards [22]. Besides the advanced features, time-travel debuggers also always provide common debugging features that are known from step-wise debuggers.

The general procedure for using time-travel debuggers is very similar to the one required for step-wise debuggers. However, depending on the implementation of the time-travel debugger, programmers have to take some setup steps. Often, the time-travel feature depends on a previous recording of the program execution, either a complete trace or a set of checkpoints of the complete run-time state. To take that recording, programmers often have to execute the program in a specialized run-time environment, such as an adapted VM. Programmers have to wait for the execution to finish before they can start debugging. Recording the execution can increase the execution duration by a factor of 7 to 300, depending on the instrumentation technique and whether a full trace or only checkpoints are recorded [22, 36]. At the same time, programmers do not have to restart the program during debugging whenever they want to go backward in time, but only when they change the source code or want to debug using different input data. Time-travel debuggers are most often standalone tools. When they are integrated into an ide, the integration results from a special plugin and not because they are a standard perspective of the IDE.

Time-travel debuggers are less common than step-wise debuggers. Nevertheless, companies and research groups have created a variety of time-travel debuggers [33, 36]. The *ZStep* debugger is an early instance of a time-travel debugger available for Lisp [24]. Programmers using it could already step backward through the source code, as it retained a history of the results of previous evaluations. The *Omniscient Debugger* records all method sends, changes to variables, and exceptions of a Java program (Figure 12) [22]. The user interface design of the Omniscient Debugger looks similar to a common step-wise debugger but highlights changes so that programmers should directly see what has changed from one step to another [22, page 227]. Several time-travel debuggers offer features beyond basic backward stepping. The *PathView* debugger integrates time-travel debugging with techniques to localize faults based on test runs [33]. Based on test runs, the tool shows a trace view that highlights the probability that the method participated in the defect. With some time-travel debuggers, programmers can navigate the trace along causal relations [36]. For instance, the *Whyline* debugging tool allows programmers to ask questions like why a statement was executed or why a variable has the present value [16]. The Whyline answers these questions by generating a dynamic slice. Similarly, the *Trace-oriented Debugger* combines causal navigation for state with an overview of the trace [36]. To collect a detailed trace in a reasonable time, the Trace-oriented Debugger has a complex architecture for collecting the run-time trace [37]. Programmers get very detailed information and causal navigation at the cost of setting up a complex infrastructure and considerable compute resources [33]. Finally, the *Undo* debugger is one of a few commercially available time-travel debuggers. To achieve low run-time overhead the underlying infrastructure, takes periodic snapshots as checkpoints for going back in time [36].

## 8 Conclusion

For live programming tools, we identified the local perspective on run-time behavior as the source of the strength of these tools and also a limiting factor when applying them to more extensive and complex programs in particular during debugging. To mitigate this, we propose to add a cross-cutting perspective on run-time behavior in combination with the local perspective. We described a prototype environment that offers three live views of a call trace and integrates feedback on control flow with probes. Participants of our exploratory user regarded the local perspective as the most useful feature and the cross-cutting views as an additional, useful perspective. Further, according to our study, the proposed environment can extend the live programming experience to debugging and working with more extensive and complex programs.

**Acknowledgements** We sincerely thank the anonymous reviewers for their detailed and valuable feedback. This work was supported by Deutsche Forschungsgemeinschaft

(DFG) grant #449591262 and the HPI–MIT "Designing for Sustainability" research program.[6]

[6] https://hpi.de/en/research/cooperations-partners/research-program-designing-for-sustainability.html (accessed February 20, 2024).

## About the authors

**Patrick Rein** is a member of the Software Architecture Group of the Hasso Plattner Institute at the University of Potsdam. He is working on improving the usability of programming tools, so programmers can use the tools most useful to them. His current research interests include the design and impact of live and exploratory programming systems. Contact Patrick at patrick.rein@hpi.uni-potsdam.de.
https://orcid.org/0000-0001-9454-8381

**Christian Flach** is a graduate student interested in trace-based, live programming tools. Contact Christian at christian.flach@hpi.uni-potsdam.de.
https://orcid.org/0000-0003-0549-5084

**Stefan Ramson** is a member of the Software Architecture Group of the Hasso Plattner Institute at the University of Potsdam. He regards the design of programming systems as the intersection of notation, interface design, psychology, and ergonomics. His current research interests include live and exploratory programming systems, alternative input methods, visual languages, and natural programming. Contact Stefan at stefan.ramson@hpi.uni-potsdam.de.
https://orcid.org/0000-0002-0913-1264

**Eva Krebs** is a member of the Software Architecture Group of the Hasso Plattner Institute at the University of Potsdam. Her research interests include example-based programming systems such as Babylonian Programming as well as using live programming systems, gamification, and educational games for computer science education. Contact Eva at eva.krebs@hpi.uni-potsdam.de.
https://orcid.org/0000-0002-9089-7784

**Robert Hirschfeld** leads the Software Architecture Group at the Hasso Plattner Institute at the University of Potsdam. His research interests include dynamic programming languages, development tools, and runtime environments to make live, exploratory programming more approachable. Contact Robert at robert.hirschfeld@hpi.uni-potsdam.de.
https://orcid.org/0000-0002-4249-6003